\documentclass[
12pt]{article}
\usepackage{color}
\usepackage{graphicx}
 \usepackage{amsmath,amssymb,array,calc}

\definecolor{darkgreen}{rgb}{0,0.55,0}

\newcommand{\bea}{\begin{eqnarray}}
\newcommand{\eea}{\end{eqnarray}}
\newcommand{\be}{\begin{equation}}
\newcommand{\ee}{\end{equation}}

\def\revise#1 {\raisebox{-0em}{\rule{3pt}{1em}}%
                     \marginpar{\raisebox{.5em}{\vrule width3pt\
                     \vrule width0pt height 0pt depth0.5em
                     \hbox to 0cm{\hspace{0cm}{%
                     \parbox[t]{4em}{\raggedright\footnotesize{#1}}}\hss}}}}

\def\sqr#1#2{{\vcenter{\vbox{\hrule height.#2pt
 \hbox{\vrule width.#2pt height#1pt \kern#1pt
 \vrule width.#2pt}\hrule height.#2pt}}}}

\catcode`\@=12

\begin{document}

\makeatletter \@addtoreset{equation}{section} \makeatother
\renewcommand{\theequation}{\thesection.\arabic{equation}}

\renewcommand\baselinestretch{1.25}
\setlength{\paperheight}{10in}
\setlength{\paperwidth}{9.5in}

\begin{titlepage}


%

\vskip 1cm

\vskip 3cm

 \begin{center}
{\bf \large All 4-dimensional static,
spherically symmetric, 2-charge abelian Kaluza-Klein black holes
and their CFT duals}
\end{center}
\vskip 1cm

\centerline{\large Edwin Barnes, Diana Vaman, Chaolun Wu }

\vskip .5cm

\centerline{\it Department of Physics, The University of Virginia}
\centerline{\it McCormick Rd, Charlottesville, VA 22904}

\centerline{Email addresses: eb4df, dv3h, cw2an@virginia.edu}

\vspace{1cm}

\begin{abstract}
We derive the dual CFT Virasoro algebras from the algebra of conserved diffeomorphism charges, for a large class of
abelian Kaluza-Klein black holes. Under certain conditions, such as non-vanishing electric and magnetic monopole charges,
the Kaluza-Klein black holes have a Reissner-Nordstrom space-time structure.
For the non-extremal charged Kaluza-Klein black holes, we use
the uplifted 6d pure gravity solutions to construct a set of Killing horizon preserving diffeomorphisms.
For the (non-supersymmetric) extremal black holes, we take the NENH limit, and construct a one-parameter family of diffeomorphisms which
preserve the Hamiltonian constraints at spatial infinity.
In each case we evaluate the algebra of conserved diffeomorphism charges following Barnich, Brandt and Compere, who used a cohomological approach, and Silva, who employed a covariant-Lagrangian formalism.
 At the Killing horizon, it is only Silva's algebra which acquires a central charge extension, and which
enables us to recover the Bekenstein-Hawking black hole entropy from the Cardy formula. For the NENH geometry, the
extremal black hole entropy is obtained only when the free parameter of the diffeomorphism generating vector fields is chosen
such that the central terms of the two algebras are in agreement. 
\end{abstract}

\end{titlepage}



\section{Introduction and Summary}

Since the proposal of a duality between rotating black holes in
four dimensions (4d) and a two-dimensional conformal field theory
\cite{Guica:2008mu}, much work has been done to generalize this correspondence
to other types of black holes. So far, the correspondence has been
extended to charged and rotating black holes in AdS in various dimensions \cite{Lu:2008jk, Hartman:2008pb}, five-dimensional (5d) KK black holes \cite{Azeyanagi:2008kb, Garousi:2009zx}, and various types of black holes in extended gravity theories and string theory \cite{Chow:2008dp, Isono:2008kx, Azeyanagi:2008dk, Chen:2009xja, Ghezelbash:2009gf, Astefanesei:2009sh}. A large class of 4d and 5d black holes with electric and scalar charges was also considered in \cite{Compere:2009dp}. We will propose that
this correspondence can be extended still further to include all 4d, static, spherically symmetric Kaluza-Klein black holes with
two sets of abelian electric and magnetic charges.

An analysis of this class of black hole solutions to 4d Einstein-Maxwell theory is facilitated by recasting them as solutions of
pure six-dimensional gravity \cite{Cvetic:1995sz, Cvetic:1995py, Cvetic:1994hv}. Depending on
the U(1) charges, these black holes have the same global space-time
structure as Schwarzschild black holes or Reissner-Nordstrom black
holes. In the BPS limit, these become supersymmetric extreme black
holes, with a null singularity. Unlike the 4d solutions, the 6d
solutions do not reside in flat space. This is a consequence of the non-vanishing 4d magnetic monopole charges.

The first bit of evidence for the existence of a CFT dual to a
given black hole arises by showing that one can construct a
classical Virasoro ($Diff(S^1)$) algebra from the Lie algebra of
 an appropriate set of
diffeomorphisms of the black hole geometry:
\be
i [\xi_m,\xi_n]_{Lie}=(m-n)\xi_{m+n}\label{diffs1}.
\ee

For non-extremal black
holes, such a Virasoro can be found by considering
diffeomorphisms which satisfy certain constraints at either the
event horizon or spatial infinity. The former approach was pioneered
by Carlip \cite{Carlip:1998wz, Carlip:1999cy, Park:2001zn}.

The analysis at spatial infinity was
first developed by Brown and Henneaux \cite{Brown:1986nw}.

For extremal black holes,
one first takes a particular near-extremal, near-horizon (NENH)
limit and imposes boundary conditions on the various fields in
the resulting geometry to obtain a set of diffeomorphisms which
obey (\ref{diffs1}) at spatial infinity.
This approach was first taken by \cite{Guica:2008mu}
and closely follows that of Brown and Henneaux.

In this paper, we
will be interested in the near-horizon analysis of non-extremal
black holes as well as the analysis instigated by \cite{Guica:2008mu}
for extremal black holes. As far as we are aware, the non-extremal
horizon analysis presented here is the first implementation of this
approach for 4d charged black holes. The novelty of our treatment is that
we can extract the Virasoro algebra of the CFT directly from the 6d
geometry where the black holes have only angular momentum and no electric or
magnetic charges, enabling us to sidestep the necessity to extend the non-extremal
horizon analysis to include electric and magnetic fields. One hopes that this approach will facilitate a near Killing horizon analysis directly for the charged 4d black holes.

For either extremal or non-extremal black holes, the next step
toward the dual CFT is to promote the classical Virasoro algebra
of diffeomorphisms to a quantum Virasoro algebra of associated
conserved charges.
The diffeomorphism charges ${\cal Q}_n$ associated with the vector fields $\xi_n$ satisfy a Poisson bracket algebra
\be
[{\cal Q}_m,{\cal Q}_n]_{P}\equiv \delta_{\xi_m} {\cal Q}_n,
\ee
which is isomorphic to the Lie algebra of the diffeomorphism generating vector fields, up to central charges
\be
i[{\cal Q}_{m},{\cal Q}_{n}]_{P}=(m-n){\cal Q}_{{m+n}}+\frac{c}{12} m(m^2-1)\delta_{m+n} \label{virasoro}.
\ee
When the central charges are non-vanishing, the classical Virasoro algebra gets promoted to a quantum Virasoro algebra, which is then used to characterize the dual CFT.

In this vein, the literature provides more than
one option. One of the most widely used diffeomorphism charge algebras was developed by Barnich, Brandt and Comp\`{e}re (BBC) \cite{Barnich:2001jy, Barnich:2007bf} from a cohomological approach:
\bea
&&[{\cal Q}_m,{\cal Q}_n]\bigg|_{central \;term} =- \frac{1}{16\pi G} \int\epsilon_{\mu_1\mu_2\dots\mu_n}
\bigg(\xi_m^{\mu_2}\nabla^{\mu_1} h -\zeta^{\mu_2}\nabla_\nu h^{\mu_1\nu} +
\xi_m^{\nu}\nabla^{\mu_2}h^{\mu_1}{}_\nu \nonumber\\
&&\;\;\;\;\;\;\;\;\;\;\;\;\;\;\;\;
+\frac 12 h \nabla^{\mu_2}\xi_m^{\mu_1}
-h^{\mu_2\nu}\nabla_\nu \xi_m^{\mu_1} + \frac 12 h^{\mu_2}{}_\nu (\nabla^{\mu_1}
\xi_m^\nu + \nabla^{\nu}\xi_m^{\mu_1}\bigg)\label{bb0},
\eea
where
\be
h_{\mu\nu}=\delta_{\xi_n} g_{\mu\nu}=\nabla_\mu\xi_\nu+\nabla_\nu\xi_\mu.
\ee
The applicability of this algebra requires a certain asymptotic
behavior of the metric and other fields.  In \cite{Barnich:2001jy}, only a set
of sufficient boundary conditions were discussed, namely the fluctuations should be subleading to the the background fields. (But this requirement is
already violated by most (if not all) of the boundary conditions
implemented in the NENH black hole analysis.) 
In \cite{Barnich:2007bf}, a more general analysis was performed, outside the confines of the linearized approximation. The conditions for the validity of the charge algebra, besides the usual requirements that the diffeomorphisms preserve the boundary conditions and that the charges be finite and integrable, included 
\bea
&&\int (\delta_\xi g^{\gamma\delta}) \frac{d}{d_v g^{\gamma\delta}} d_v \bigg(\frac{\sqrt{g}}{16\pi G} g_{\alpha\beta} \,d_v g^{\mu\alpha}\,d_v g^{\nu\beta} (d^{n-2} x)_{\mu\nu}\bigg)\approx o(r^0)\nonumber\\
&&\int \delta_{\xi}\bigg(\frac{\sqrt{g}}{16\pi G} g_{\alpha\beta} \,d_v g^{\mu\alpha}\,d_v g^{\nu\beta} (d^{n-2} x)_{\mu\nu}\bigg)\approx o(r^0)
\label{condition}\eea
where the integral is taken over some boundary at fixed $r,t$.
In the above $d_v$ is a field vertical differential, $d_v^2 g^{\mu\nu}=0$. The authors of \cite{Barnich:2007bf} noted that it is possible to remove the constraint (\ref{condition}) at the price of adding boundary terms to the Lagrangian.
We will comment on the importance of this condition when testing the applicability of the BBC algebra.

On the other hand, Silva \cite{Silva:2002jq,Julia:2002df} proposed an alternative construction for
the charge algebra, based on the Lagrangian Noether method, which is purported to have no dependence on
boundary conditions for the fields\footnote{The same central term was obtained by Koga \cite{Koga:2001vq} from the asymptotic symmetries at Killing horizons, by using the covariant phase space formalism.}:
\bea
&&[{\cal Q}_m,{\cal Q}_n] \bigg|_{central\;term}=\frac{1}{16\pi G} \int\epsilon_{\mu_1\mu_2\dots \mu_n}\bigg(
\nabla^\nu \xi_m^{\mu_2} \nabla_\nu \xi_n^{\mu_1}-
\nabla^\nu \xi_m^{\mu_1}\nabla_\nu \xi_n^{\mu_2} - R^{\mu_2\mu_1}{}_{\nu\rho}
\xi_m^\nu\xi_n^\rho \nonumber\\
&&\;\;\;\;\;\;\;\;\;\;\;\;\;\;\;\;
+\xi_m^{\mu_1}\xi_n^{\mu_2}(R-2\Lambda)+
\frac{1}2 (\nabla^{\mu_2}\xi_m^{\mu_1} - \nabla^{\mu_1}\xi_m^{\mu_2})\nabla_\nu \xi_n^\nu-
\frac{1}2 (\nabla^{\mu_2}\xi_n^{\mu_1} - \nabla^{\mu_1}\xi_n^{\mu_2})\nabla_\nu \xi_m^\nu\bigg).\nonumber\\
\label{silva0}
\eea
Silva noticed that, while constructing  the charges ${\cal Q}_n$ is a boundary condition-dependent issue, the charge variation $\delta {\cal Q}_n$ depends only on the equations of motion and on the gauge symmetry. This leads to the definition of the Poisson bracket of the two charges and to the central term quoted above (\ref{silva0}). Lastly, we'd like to mention that in \cite{Silva:2002jq,Julia:2002df}, the total derivative terms which can be added to the Lagrangian, leading to ambiguities in the charges, are fixed by enforcing the variational principle.

Although Silva's algebra can be rewritten in a form which closely resembles the BBC algebra, the two central terms agree only up to 
\be
\frac{1}{16\pi G}\int \epsilon_{\mu_1\mu_2\dots\mu_n}\frac 12 (\nabla^{\mu_1}\xi_m^{\nu}+\nabla^\nu\xi_m^{\mu_1})(\nabla^{\mu_2}\xi_{n\,\nu}+\nabla_\nu\xi_{n}^{\mu_2}), \label{diffterm}
\ee
 after partial integration.
According to Barnich and Brandt, as reported by Silva \cite{Silva:2002jq}, this term vanishes for all examples which were known at the time of his work.
More recently, in \cite{Barnich:2007bf}, the integrand of the difference term (\ref{diffterm}) was recognized as the $n-2$ form which enters the supplementary requirement (\ref{condition})
\be
E(d_v g, d_v g) = \frac{\sqrt{g}}{16\pi G} g_{\alpha\beta}\, d_v g^{\mu\alpha}\, d_v g^{\nu\beta} (d^{n-2} x)_{\mu\nu}
\ee
after replacing $d_v g^{\mu\nu}$ by $-\nabla^\mu \xi^\nu - \nabla^\nu \xi^\mu$. 

In addition to presenting a new class of black hole-CFT examples,
a second goal of this paper is to compare these two charge
algebras. 
We will
show that the same set of diffeomorphisms can lead to two different
values for the central charge of the CFT, depending on which charge
algebra is applied. However, we need to proceed with caution, 
since some of the conditions necessary for the validity of the algebras can be violated. 

Unfortunately, very little is known about the dual CFT in most instances
of the black hole-CFT correspondence, except in cases where the
black hole is a solution of string theory. For the most
part, the only information about the CFT that one can obtain is
the central charge and whether or not the CFT is chiral. Fortunately,
knowledge of the central charge suffices for a computation of
the CFT entropy from either the usual Cardy formula or its thermal
version.

We can exploit Cardy's relation between the central charge and the
entropy $S=2\pi \sqrt{c{\cal Q}_0/6}$ to test the two different charge algebras. Since the two
algebras yield, in principle, differing central charges, they will produce
different entropies. These can then be compared with the Bekenstein-
Hawking entropy of the black hole, which is expected to coincide
with the CFT entropy. 
For non-extremal black holes, we
use the uplifted 6d solutions and follow Carlip's approach by 
first identifying a set of diffeomorphisms which preserve the horizon structure. Then, for that same set of diffeomorphisms, we compute the central term of the BBC and Silva's algebras. We find that the BBC algebra does not acquire a central term extension. However, it turns out that (\ref{condition}) is violated, and the BBC algebra is simply not applicable. On the other hand, Silva's algebra aquires a central extension, and through the use of Cardy's formula yields
the Bekenstein-Hawking result for the balck hole entropy.  
The zero-mode ${\cal Q}_0$ is obtained from the Poisson bracket algebra, by requiring that we cast it in the standard Virasoro form (\ref{virasoro}).

For extremal black holes, the situation is more complicated. The class of Kaluza-Klein black holes which we study yield the same $AdS_2\times S^2\times T_2$ geometry. From a 4d perspective, the geometry is $AdS_2\times S^2$, supported by non-vanishing electric and magnetic Maxwell fields, and scalar fields. For simplicity, rather than dealing directly with the 4d NENH limit, we use the 6d uplifted geometry. In
this case, we need to impose boundary conditions on the metric at
spatial infinity of the NENH geometry. We were able to find a one-parameter family of boundary conditions which obey all the known
linear self-consistency constraints. These boundary conditions in
turn lead to a one-parameter family of diffeomorphisms. The
arbitrary parameter carries over to the diffeomorphism charge algebra,
and appears in the expression for the central charge regardless of
which charge algebra is used\footnote{Although Silva's algebra is
advertised as being independent of boundary conditions, there is
no obvious way to select a set of diffeomorphisms by fixing the
free parameter from the field boundary conditions alone.  }${^,}$\footnote{ We note that the validity conditions for the BBC algebra, and in particular (\ref{condition}), are responsible for separately fixing the value of this free parameter. We are grateful to G. Comp\`{e}re for bringing to our attention the supplementary constraint (\ref{condition}).}
. We further observe that the term describing
the difference between the two algebras depends on the free parameter.
When the parameter is chosen to be such that this difference term
vanishes, then both algebras lead to the Bekenstein-Hawking entropy. We determine the temperature of the dual CFT by expressing the extremal black hole entropy in terms of quantized charges $dS_{extremal}=\sum_i \beta_i dQ_i$.
Using the central charge value common to both Silva and BBC charge algebras, we were able to recover from the thermal Cardy formula $S=\pi^2 c/(3 \beta)$ the entropy of the extremal Kaluza-Klein black holes.

The paper is organized as follows. In section 2, we briefly review
the work of Cvetic and Youm and tailor it to arrive at a class of
6d solutions which compactify to the class of 4d multi-charged
solutions described earlier. In the same section, we also compute
the 4d charges along with other thermodynamic quantities that will
be relevant for later sections. In section 3, we construct the
geometry of the NENH limit which will be necessary for our discussion
of extremal black holes.  Section 4 focuses on obtaining the CFT
entropy in the non-extremal case from a near Killing horizon analysis.
Section 5
is devoted to deriving the extremal black hole entropy from the asymptotic algebra(s) of the diffeomorphism charges.

\section{All 4-dimensional static, spherically symmetric, \\2-charge, abelian Kaluza-Klein black holes}

\subsection{Geometry}

Cvetic and Youm constructed a large class of four-dimensional static, spherically symmetric black holes of (4+n)-dimensional abelian Kaluza-Klein theory. Their construction was based on the observation that for such solutions, one has a (n+1)-parameter abelian isometry group (time-translation being one of the isometries). Then the solutions of (4+n)-dimensional pure gravity,
\be
g_{\Lambda\Pi}=
\begin{pmatrix}
\exp(-\frac \varphi\alpha ) g_{\mu\nu}+\exp(\frac{2\varphi}{n\alpha})\rho_{ij}
A_\mu^i A_\nu^j & \exp(\frac{2\varphi}{n\alpha})\rho_{ij} A^j_\lambda\\
\exp(\frac{2\varphi}{n\alpha})\rho_{ij}A^i_\pi &\exp(\frac{2\varphi}{n\alpha})\rho_{ij}
\end{pmatrix}\label{KK},
\ee
where $g_{\mu\nu}$ is the 4d metric, $A_\mu^i$ are the $n$ U(1) gauge fields, $\rho_{ij}$ is the unimodular internal metric and $\alpha = \sqrt{n+2}/\sqrt{n}$,
are obtained from the effective three dimensional (3d) action \cite{Dobiasch:1981vh}:
\be
S=-\frac 12\int \sqrt{g_{3}} \bigg(R_3 +\frac 12 Tr(\chi^{-1}\partial_a\chi \chi^{-1}\partial^a \chi)\bigg)\label{effaction},
\ee
where $g_{3}=\det(\tau g_{ab}^\perp), a,b =1,2,3$, $R_3$ is the Ricci scalar associated with the effective three-dimensional metric $\tau g_{ab}^\perp$, and
\be
\chi=\begin{pmatrix}
\tau^{-1}&-\tau^{-1}\omega^T
\\
-\tau^{-1}\omega&\check\lambda + \tau^{-1}\omega\omega^T
\end{pmatrix}.
\ee
The relationship between the 4+n-dimensional metric and the effective action fields is as follows. Given the (n+1) Killing vectors $k_I$, the metric $g_{\Lambda\Pi}$ decomposes into
\bea
&&\check\lambda_{IJ}=g_{\Lambda\Pi}k_{I}^\Lambda k_J^\Pi\\
&&g^\perp_{\Lambda\Pi}=g_{\Lambda\Pi}-\lambda^{IJ}k_{I\Lambda}k_{J\Pi}\\
&&g_{\Lambda\Pi}=\begin{pmatrix}
g_{ab}^\perp + \lambda^{IJ} k_{Ia}k_{Jb} & k_{Ka}\\
k_{Jb}&\check\lambda_{JK}
\end{pmatrix}, \qquad a,b=1,2,3; I,J,K=1,\dots n+1.
\eea
Furthermore, for Einstein spaces $R_{\Lambda\Pi}=0$,
\be
\omega_{Ia}= \epsilon_{abc}k_{I}^{b;c}=\partial_a\omega_I.
\ee
Lastly, with
\be
\tau=\det(\check\lambda),
\ee
it can be shown \cite{Maison:1979kx, Dobiasch:1981vh} that the (4+n)-dimensional pure gravity equations of motion $R_{\Lambda\Pi}=0$ reduce to those of the effective action (\ref{effaction}). Notice that $\chi$ is a unimodular matrix, which for 4d spherically symmetric solutions depends only on the radial coordinate $r$ of the 4d subspace. Also, for such solutions the effective 3d metric takes the form
\be
\tau g_{ab}^\perp = diag(1,f(r),f(r)\sin^2\theta).
\ee
The 4d metric, gauge and scalar fields in terms of the effective 3d fields are given by
\bea
&&\exp(-\frac \varphi \alpha)g_{\mu\nu}=diag(-\tau^{-1}, -\tau^{-1}f(r),-\tau^{-1}f(r)\sin\theta,(\check\lambda^{11})^{-1})\\
&&\exp(-\frac{2\varphi}{n\alpha})\rho_{IJ}=\check\lambda_{I+1,J+1}\\
&&A_0^I=-\frac{\check\lambda^{I+1}}{\check\lambda^{11}},
\qquad A_\phi^I=-\tau^{-1}\exp(\frac{2\varphi}{n\alpha})\rho^{IJ}f(r)\frac{d}{dr}\omega_{J+1}\cos\theta,
\eea
where the parameter $\alpha$ is chosen such that the dilaton $\varphi$ has a standard kinetic term: $\alpha=\sqrt{n+2}/\sqrt{n}$.

Cvetic and Youm employed a solution-generating technique, which allowed them to obtain all 4d static, spherically symmetric, abelian Kaluza-Klein solutions, by starting from the 4d Schwarzschild black hole solution
\bea
&&f(r)=r(r-2\gamma)\\
&&\chi=diag(-(1-\frac {2\gamma}r)^{-1}, -(1-\frac {2\gamma}r), 1,\dots 1),
\eea
and acting with $SO(2,n)$ transformations.

The solutions which we are interested in, for the purpose of this paper, have $n=2$, and arise through the following succesion of transformations. Start from
\bea
&&f(r)=r(r-2\gamma)\\
&&\chi=diag(-(1-\frac {2\gamma}r)^{-1}, -(1-\frac {2\gamma}r), 1,1),
\eea
and first perform an SO(1,1) rotation on $\chi$ which acts on the 1st and 3rd indices, with parameter $\delta_P$, and another SO(1,1) rotation acting on the 2nd and 4th indices, with parameter $\delta_Q$.
Defining
\bea
&&P=2\gamma\sinh(\delta_P)\cosh(\delta_P)\equiv\sqrt{p(p-2\gamma)},\\
 &&Q=2\gamma\sinh(\delta_Q)\cosh(\delta_Q)\equiv\sqrt{q(q-2\gamma)},
\eea
the rotated $\chi$ matrix reads
\be
\chi=\begin{pmatrix}-\frac{r+p}r&0&\frac{P}r&0\\
0&-\frac{r+2\gamma-q}{r+2\gamma}&0&\frac{Q}{r+2\gamma}\\
\frac Pr&0&\frac{r+2\gamma-p}r&0\\
0&\frac{Q}{r+2\gamma}&0&\frac{r+q}{r+2\gamma}
\end{pmatrix},
\ee
and $f(r)=r(r+2\gamma)$. In the above we have also redefined $r$ by sending $r-2\gamma$ into $r$.
The solutions generated in this fashion have the same global space-time structure as 4d Schwarzschild black holes.
To arrive at more general ones, with an inner and outer horizon, we continue the set of SO(2,2) rotations, with an SO(1,1) transformation with parameter $\delta_1$, mixing the 1st and 4th indices, followed by yet another SO(1,1) rotation, mixing the 2nd and 3rd indices, with parameter $\delta_2$, where
\bea
&&\delta_1=-{\rm{arctanh}}(\sqrt{p(p-2\gamma)}\delta),\nonumber\\
&&\delta_2={\rm{arctanh}}(\sqrt{q(q-2\gamma)}\delta)\label{d1d2}.
\eea
The parameters $\delta_1, \delta_2$ are not independent. The condition that the metric obtained after these transformations asymptotes to flat space, and thus has a vanishing Taub-NUT charge restricts $\delta_1$ and $\delta_2$ as in (\ref{d1d2}).

As advertised, these 4d black holes are solutions of the Kaluza-Klein reduction of 6d Einstein gravity. The reduced action is obtained by substituting the Kaluza-Klein ansatz (\ref{KK}) into the Einstein-Hilbert action
\be
{\cal S}=\frac{1}{16\pi G} \int d^4 x \sqrt{g} (R+\frac 14 \rho_{ij} e^{\alpha\varphi} F_{\mu\nu}^{i}F^{\mu\nu;j}-\frac 12 \partial_\mu \varphi\partial^\mu\varphi-\frac 14
\rho^{ij}\rho^{kl}\partial_\mu \rho_{ik}\partial^\mu\rho_{jl})\label{KKaction},
\ee
where we recall that $\alpha=\sqrt{2}$ and $\rho$ is a unimodular $2\times 2$ matrix. The corresponding 6d uplift is characterized by
\bea
&&ds_6^2=-N^2 dt^2+g_{rr} dr^2+g_{\theta\theta}(d\theta^2+\sin^2\theta d\phi^2)+g_{yy}(dy + A_0^y d t +A_\phi^y d\phi)^2\nonumber\\
&&\;\;\;\;\;\;+g_{zz}(dz +A_0^z dt+A_\phi^z d\phi)^2
+2g_{yz}(dy+A_0^y dt+A_\phi^y d\phi)(dz+A_0^z dt+A_\phi^z d\phi)\nonumber\\
&&\tau=-\frac{(-1+p^2\delta^2-2p\delta^2\gamma)r(r+2\gamma)}{(-r^2-2\gamma r-pr-2\gamma p+p^2\delta^2 r^2+\delta^2 p^2 r q-2 p\delta^2\gamma r^2-2q\delta^2 p r
\gamma)}\nonumber\\
&&N^2=-\frac{(r+2\gamma)(-1+q^2\delta^2-2q\delta^2\gamma)r}{-2q\delta^2pr\gamma+q^2\delta^2 pr-r^2-rq-2q\delta^2r^2\gamma+q^2\delta^2r^2-2\gamma r-2q\gamma}\nonumber
\eea
\bea
&&g_{rr}=-\tau^{-1}\nonumber\\
&&g_{\theta\theta}=\frac{-r^2-2\gamma r-p r-2\gamma p+p^2\delta^2 r^2+\delta^2 p^2 r q-2p\delta^2\gamma r^2-2q\delta^2pr\gamma}{-1+p^2\delta^2-2p\delta^2\gamma}\nonumber\\
&&A_0^y=-\frac{(-pr+2\gamma q+2\gamma r+r q)\delta\sqrt{q(q-2\gamma)}}{(-2q\delta^2 p r\gamma+q^2\delta^2 p r+q^2\delta^2 r^2-2 q\delta^2 r^2\gamma-2 \gamma q-r^2-2\gamma r-r q)}\nonumber\\
&&A_\phi^y=\frac{\sqrt{1-q^2 \delta^2+2 q\delta^2\gamma}\sqrt{p(p-2\gamma)}}{\sqrt{1-p^2\delta^2+2p\delta^2\gamma}}\cos\theta\nonumber\\
&&A_0^z= -\frac{(-1+q^2\delta^2-2q\delta^2\gamma)(p^2\delta^2 r-2p\delta^2 r\gamma-r-2\gamma)}
{(r p\delta^2 q^2-2 p\delta^2 r\gamma q+r^2 q^2\delta^2-2\delta^2 r^2 q\gamma -r^2-2\gamma q-q r-2 r\gamma)}\nonumber\\
&&\;\;\;\;\;\;\times\frac{\sqrt{q(q-2\gamma)}}{\sqrt{1-q^2\delta^2+2 q\delta^2\gamma}\sqrt{1-p^2\delta^2+2p\delta^2\gamma}}
\nonumber\\
&&A_\phi^z=
\frac{\sqrt{p(p-2\gamma)}\delta(-2\gamma-p+q)}{-1+p^2\delta^2-2p\delta^2\gamma}\cos\theta\nonumber\\
&&g_{yy}= (r^2-p^2 \delta^2 r^2+2 p \delta^2 \gamma r^2-q^2 \delta^2 r^2+q^2 \delta^4 r^2 p^2-2 q^2 \delta^4 r^2 p \gamma+2 q \delta^2 r^2 \gamma-2 q \delta^4 r^2 \gamma p^2\nonumber\\
&&\;\;\;\;\;\;
+4 q \delta^4 r^2 \gamma^2 p+4 \gamma r+4 q \delta^2 p r \gamma
-4 \delta^2 p^2 r \gamma+p^3 \delta^2 r-q^2 \delta^2 p r+4 q \delta^2 r \gamma^2+4 p \delta^2 \gamma^2 r-4 q^2 \delta^2 r \gamma\nonumber\\
&&\;\;\;\;\;\;+q^3 \delta^2 r-\delta^2 p^2 r q+p q^3 \delta^2+\delta^2 p^3 q+4 \gamma^2-2 \delta^2 p^2 q^2)\nonumber\\
&&\;\;\;\;\;\;
\times\frac{1}{(-1+q^2 \delta^2-2 q \delta^2 \gamma)(-r^2-2 \gamma r-p r-2 \gamma p+p^2 \delta^2 r^2+\delta^2 p^2 r q-2 p \delta^2 \gamma r^2-2 q \delta^2 p r \gamma)}\nonumber\\
&&g_{zz}= \frac{(-1+p^2\delta^2-2 p\delta^2\gamma)(r+p)(r+q)}{(-r^2-2r\gamma-p
r-2p\gamma+p^2\delta^2 r^2+p^2\delta^2 r q-2 p\delta^2 r^2\gamma-2 p\delta^2 r\gamma q)}\nonumber\\
&&g_{yz}=-\frac{(-1+p^2\delta^2-2p\delta^2\gamma)(p-q)(pr+qp-2r\gamma+qr)\delta}{(-r^2-2r\gamma -p r-2 p\gamma +p^2\delta^2 r^2+p^2\delta^2 rq-2p\delta^2 r^2\gamma-2p\delta^2 r\gamma q)}\nonumber\\
&&\;\;\;\;\;\;\times\frac{1}{\sqrt{1-q^2\delta^2+2q\delta^2\gamma}\sqrt{1-p^2\delta^2+2p\delta^2\gamma}}\label{6duplift}.
\eea

{}From the zeros of the lapse function $N^2$ we infer that the position of the outer horizon is at $r_+=0$, whereas the inner horizon is at $r_-=-2\gamma$.
Therefore $\gamma$ is the non-extremality parameter, defined as $(r_+-r_-)/2$.
We should perhaps emphasize that having either one of the parameters $\delta_P, \delta_Q, \delta$ vanishing leads to solutions which do not have the same global space-time structure as 4d Reissner-Nordstrom black holes. The supersymmetric black holes are charcterized by $\gamma=0$, with $p,q$ fixed.

\subsection{Thermodynamics}

The ADM mass of the 4-dimensional Kaluza-Klein black holes is
\bea
M_{ADM;4d}&=&\frac{1}{8\pi G_4} \int dS_{\mu\nu} \partial^\mu k^\nu\nonumber\\
&=&\frac{(4\delta^2\gamma^2 -1-2p\delta^2 \gamma+q\delta^2p-2q\delta^2\gamma)(qp^2\delta^2+pq^2\delta^2-p-q-4pq\delta^2\gamma)}{4G_4(1-p^2\delta^2+2p\gamma\delta^2)(1-q^2\delta^2+2q\gamma \delta^2)},\nonumber\\
\eea
The electric and magnetic fields and the corresponding electrostatic and magnetostatic potentials are
\bea
&&E^i = \partial_r A^i_0 \equiv \partial_r \Phi^i, \qquad i,j=1,2=y,z\\
&&B_i = e^{\alpha\varphi} \rho_{ij} \partial_\theta A_\phi^j \equiv \partial_r\Psi_i.
\eea
It is amusing to note that the magnetostatic potentials are related to the electrostatic ones as follows: to obtain $\Psi_1= \Psi_y$ start from $\Phi^2 = A_0^z$ and interchange $p$ and $q$; to obtain $\Psi_2=\Psi_z$ start from $\Phi^1=A_0^y$ and interchange $p$ and $q$.

{}The conserved electric and magnetic monopole charges
\bea
&&Q_1=-\frac{(p-q-2\gamma)\sqrt{q(q-2\gamma)} \;\delta}{1-q^2\delta^2+2q\delta^2\gamma},\\
&&Q_2=\frac{\sqrt{1-p^2\delta^2+2p\delta^2\gamma}\sqrt{q(q-2\gamma)}}{\sqrt{1-q^2\delta^2+2q\delta^2\gamma}},\\
&&P^1=\frac{\sqrt{1-q^2\delta^2+2q\delta^2\gamma}\sqrt{p(p-2\gamma)}}{\sqrt{1-p^2\delta^2+2p\delta^2\gamma}},\\
&&P^2=\frac{(p-q+2\gamma)\sqrt{p(p-2\gamma)} \;\delta}{1-p^2\delta^2+2p\delta^2\gamma}.
\eea
can be obtained from the asymptotic behavior of the electrostatic and magnetostatic potentials $A^i \sim Q_i/r +{\cal O}(r^{-2}),$ $\Psi_i\sim P^i/r + {\cal O}(r^{-2})$.
The same pattern of simultaneously interchanging the indices 1 and 2 and the parameters $p$ and $q$ relates the magnetic and electric charges.

The Hawking temperature and horizon area are
\bea
T_{H}=\frac{\sqrt{(1-p^2\delta^2+2p\delta^2\gamma)(1-q^2\delta^2+2q\delta^2\gamma)}}{4\pi \sqrt{pq}},\\
A_{horizon;4d}=\frac{8\pi\beta\sqrt{pq}}{\sqrt{(1-p^2\delta^2+2p\delta^2\gamma)(1-q^2\delta^2+2q\delta^2\gamma)}}.
\eea

Finally, for the 4d black holes, the Smarr formula reads
\be
M_{ADM;4d}=\frac 1{2G_4} T_H A_{horizon;4} + \frac 1{4G_4}(Q_i \Phi^i + P^i \Psi_i),
\ee
and we can explicitly verify that it is satisfied.
The normalization of the Maxwell action in (\ref{KKaction}) is responsible for the peculiar factor of $1/4$ in the Smarr formula.

The extremal limit is a bit subtle. As the non-extremality parameter $\gamma$ is taken to zero,
we must also take $p\to q$, $\delta \to 1/q$:
\bea
&&p=q+(C^2-D^2)\gamma, \qquad \gamma\to 0
\nonumber\\
&&\delta=\frac{1}{q}-\frac{(C^2-1)}{q^2}\gamma, \qquad \gamma \to 0.
\eea
while keeping $C,D$ fixed.
This will insure that the lapse function has a degenerate double root at $r=0$, as expected for the extremal limit of black holes with Reissner-Nordstrom global space-time structure. The Hawking temperature vanishes for these extremal black holes.

In the extremal limit, the Bekenstein-Hawking formula for the black hole entropy yields:
\be
S_{extremal;4d} =\frac{ \pi q^2}{C D G_4}\label{S_extremal_4}.
\ee
It is interesting to notice that we can express the extremal entropy solely in terms of charges
\bea
&&S_{extremal;4d}= \frac{\pi}{2G_4}
(Q_1 P^1 + Q_2 P^2)\bigg|_{extremal}\nonumber\\
&&Q_{1\,extremal}=-\frac{q(C^2-D^2-2)}{2C^2},\qquad Q_{2\,extremal}=\frac{qD}{C}\nonumber\\
&&P^1_{extremal}=\frac {Cq}{D}, \qquad P^2_{extremal}= \frac{q(C^2-D^2+2)}{2D^2}.
\eea

The uplifted 6d solutions are non-asymptotically flat rotating black holes. The electric charges of the 4d black holes are mapped into conserved angular-momentum type charges, obtained via Komar integrals
\bea
&&J_i = \frac{1}{16\pi G_6} \int dS_{\mu\nu} \nabla^\mu l_i^\nu, \qquad i=1,2=y,z\nonumber\\
&& l_1= l_1^\mu \partial_\mu = \partial_y, \qquad l_2 = l_2^\mu \partial_\mu = \partial_z.
\eea
The magnetic monopole charges are mapped into the periodicities of the $y,z$ coordinates. In order to avoid singularities at $\theta=0,\pi$, $y$ and $z$ must be periodic with periods
\be
y \sim y+L_y=y+ 4\pi P^1, \qquad z \sim z+L_z = z+ 4\pi P^2.
\ee
The 4d Newton's constant is expressed in terms of the 6d Newton's constant as
\be
\frac{1}{G_4}= \frac{L_y L_z}{G_6},
\ee
and the Komar angular momenta can be written as
\be
J_i = \frac{Q_i}{4G_4}.
\ee
The Hawking temperature of the uplifted 6d black holes is the same as that of the 4d black holes.

In the extremal limit, the 6d black hole entropy equals
\bea
&&S_{extremal;6d}=S_{extremal;4d}\frac{G_4 L_{y\,extremal}L_{z\,extremal}}{G_6}\nonumber\\
&&\;\;\;\;\;\;=\frac{8\pi^3 q^4 (C^2-D^2+2)}{D^4 G_6}\label{S_extremal}.
\eea


\section{The NENH limit}

For later purposes we record here the near-extremal near horizon (NENH) limit of these black holes.
We begin by making a coordinate transformation
\be
r=2\epsilon\zeta,
\ee
while at the same time taking the non-extremality parameter to zero
\be
\gamma = \beta \epsilon^2,
\ee
as $\epsilon \to 0$.
As the extremality limit is approached, we need to take $p\to q$ , and $\delta\to 1/q$ as follows:
\be
p=q+\beta \epsilon^2 (C^2-D^2)\,\qquad
\delta=\frac{1}{q} - \beta\epsilon^2 \frac{C^2-1}{q^2},
\ee
where $q, C, D$ are held fixed.

To leading order in $\epsilon$, the parameter $\beta$ drops out, and the NENH geometry is $AdS_2\times S^2\times T_2$:
\bea
&&ds_6^2=-\frac{4 C^2 \zeta^2 }{q^2} dt^2 +\frac{q^2}{D^2 \zeta^2} d\zeta^2 + \frac{q^2}{D^2} d\Omega_2^2\nonumber\\
&&\;\;\;\;\;\;+
\frac{4D^2}{(b (C^2-D^2)+2 CD)^2+4} \bigg( d\tilde y -\frac{b(C^2-D^2+2)+2 CD}{Dq}\zeta dt\nonumber\\
&&\;\;\;\;\;\;+
\frac{b(C^2-D^2+2)+2 CD}{2D^2} q \cos\theta d\phi\bigg)^2\nonumber\\
&&\;\;\;\;\;\;
+\frac{D^2 (b (C^2-D^2)+2CD)^2}{(b(C^2-D^2)^2+2CD)^2+4}\bigg(d\tilde z-\frac{2C(b(C^2-D^2-2)+2CD)}{qD(b(C^2-D^2)+2CD} \zeta dt \nonumber\\
&&\;\;\;\;\;\;\;
+ \frac{b(C^2-D^2-2)+2CD}{D^2(b(C^2-D^2)+2CD}q\cos\theta d\phi\bigg)^2,\nonumber\\
\eea
where we have redefined the time coordinate $$t \to t/ \epsilon.$$ The Kaluza-Klein coordinates $(y,z)$ have been also redefined as
$$(\tilde y, \tilde z) = (y+b z, z -\frac{b((C^2-D^2)^2 +4)+ 2CD(C^2-D^2)}{2CD(b(C^2-D^2)+2CD)}y) $$
in order to diagonalize the $T_2$ line element.
The metric diagonalization can be achieved with an arbitrary parameter $b$.

A few comments are in order: the $AdS_2\times S^2$ factor is to be expected since the extremal limit of the 4d Kaluza-Klein black holes, which have the same global space-time structure as Reissner-Nordstrom black holes, is indeed $AdS_2\times S^2$. The electric and magnetic fluxes which support the $AdS_2\times S^2$ geometry are obtained by performing a straightforward Kaluza-Klein reduction along $T_2$. Second, we notice in the NENH limit the magnetic fluxes cannot be simultaneously zero. This will bring an element of novelty in constructing the diffeomorphism algebra at spatial infinity.

To simplify our analysis of the CFT dual to the 6d NENH geometry, we will further choose
$$b=-\frac{2CD}{C^2-D^2+2}.$$
With this choice, and with one more obvious rescaling of the $t, y, z $ coordinates,
$$(t,\tilde z) \to (\frac {D}{2C} t, \frac{\sqrt{2}}D \tilde z), $$
the NENH metric takes the simple form
\bea
&&ds_6^2=-\frac{D^2 \zeta^2}{q^2} dt^2 +\frac{q^2}{D^2 \zeta^2} d\zeta^2 + \frac{q^2}{D^2} d\Omega_2^2
+ (d\tilde z - \frac{\sqrt{2}D}q\zeta dt + \frac{\sqrt{2}q}{D}\cos\theta d\phi)^2 + d\tilde{\tilde y}^2.
\nonumber\\
\label{6ddiagonal}
\eea
The coordinate $\tilde z$ must be periodic with period $4\pi \sqrt{2}q/D$, to avoid any singularities.
The orginal $y,z$ coordinates, in the NENH limit, are periodic with periods
\be
y\sim y+ 4\pi\frac{Cq}{D}, \qquad z\sim z+2\pi \frac{q(C^2-D^2+2)}{D^2}.
\ee

\section{Entropy of the non-extremal black hole from symmetries at the Killing horizon}

We will follow Carlip's approach \cite{Carlip:1998wz, Carlip:1999cy}to determining the entropy of the 4d 2-charge non-extremal Kaluza-Klein black holes from symmtries at the Killing horizon. However, as advertised, we will instead work with the uplifted 6d solutions (\ref{6duplift}), which lend themselves to a more direct application of Carlip's ideas. The first step is to determine a set of differomorphisms which preserve the structure of the horizon, and which generate through their Lie brackets a $Diff(S^1)$ algebra.  A  choice satisfying these requirements is 
\be
\xi_n^\mu=T_n\chi^\mu + R_n\rho^\mu,\label{Khdiffeo}
\ee
where $\chi^\mu$ is the Killing vector whose norm vanishes at the horizon
\bea
&&\chi=\chi^\mu\partial_\mu= \partial_t +\Omega_y\partial_y + \Omega_z
\partial_z,\nonumber\\
&&\Omega_y=-A_0^y\bigg|_{r=0},\qquad \Omega_z=-A_0^z\bigg|_{r=0},
\eea
and $\Omega_y,\Omega_z$ are angular velocities at the horizon. The vector $\rho$ is given by
\be
\rho_\mu=-\frac{1}{2\kappa}\partial_\mu (\chi^\nu\chi_\nu),
\ee
where $\kappa$ is, as usual, the surface gravity:
\be
T_H=\kappa/(2\pi).
\ee
Both $\rho$ and $\chi$ have vanishing norms at the horizon.
The scalar functions $T_n$ and $R_n$ are related to each other
\be
R_n=-\frac{1}{\kappa}\chi^\mu \partial_\mu T_n + {\cal O}(r).
\ee
Our choice for $T_n$ is
\be
T_n=\frac{1}{\alpha+\Omega_y (2\pi)/L_y }e^{in(\alpha t + (2\pi)/L_y\, y)},\qquad n\in {\bf Z}.
\ee
The normalization factor for $T_n$ is such that $\xi_n$ will generate a properly normalized $Diff(S^1)$ algebra. A different choice of the $U(1)$ direction selected to write $T_n$ will not change any of our conclusions.

Next, instead of the algebra of the conserved diffeomorphism charges used by Carlip, we will employ the corrected algebra of Koga and Silva (\ref{silva0}). The central term of this algebra equals
\be
-i\delta_{m+n}m^3\frac{\alpha+\Omega_y (2\pi)/L_y}{\kappa}\frac{Area_{horizon;6d}}{G_6}=-i\delta_{m+n}m^3\frac{\alpha+\Omega_y (2\pi)/L_y}{\kappa}\frac{Area_{horizon;4d}}{G_4}.
\ee
Requiring that the central term has the standard form of a Virasoro algebra determines uniquely the zero mode ${\cal Q}_0$
\be
{\cal Q}_0=\frac{\alpha+\Omega_y (2\pi)/L_y}{\kappa}\frac{Area_{horizon;6d}}{16\pi G_6},
\ee
and the central charge of the chiral Virasoro algebra
\be
c=\frac{\alpha+\Omega_y (2\pi)/L_y}{\kappa}\frac{3 Area_{horizon;6d}}{2\pi G_6}.
\ee
Lastly, the Cardy formula leads to the Bekenstein-Hawking entropy
\be
S=2\pi \sqrt{\frac{c{\cal Q}_0}{6}}= \frac 14\frac{Area_{horizon;6d}}{G_6}=\frac 14\frac{Area_{horizon;4d}}{G_4},
\ee
provided that
\be
\alpha=\kappa-\Omega_y (2\pi)/L_y. \label{choice}
\ee
This is perhaps in accord to expectations, since the coordinates of a zero angular momentum coordinate system at the horizon are $Y=y-\Omega_y t, Z=z-\Omega_z t$. In addition, the period of the $t$ coordinate is $2\pi/\kappa$, which means that the Fourier mode expansion at the horizon will involve the scalar functions $\exp(i n (\kappa t + (2\pi)/ L_y Y))$.

When using the BBC Poisson bracket algebra of the conserved diffeomorphism charges, we find that the central term is vanishing\footnote{In fact, the constraint (\ref{condition}) is not satisfied for the diffeomorphism generator (\ref{Khdiffeo}), which means that the BBC algebra cannot be used in this case.  }. Thus for the same set of Killing horizon preserving diffeomorphisms (\ref{Khdiffeo}), only Silva's charge algebra acquires a central charge extension. Moreover, in the latter case, with the natural choice (\ref{choice}), the diffeomorphism charges yield a dual CFT Virasoro algebra which correctly accounts for the black hole entropy.

\section{Entropy of the extremal black hole from symmetries of the NENH geometry at spatial infinity}

In this section we will consider the diagonalized 6d NENH geometry (\ref{6ddiagonal}), where we notice that the $\tilde y$ coordinate parametrizes a trivial $S^1$ fibration. Thus the geometry is effectively five dimensional. The boundary conditions we impose on the metric fluctuations at the boundary of the NENH geometry, at $1/\zeta=0$, are as follows
\bea
\delta g_{\mu\nu}=
\begin{pmatrix}
 \zeta^2 h_{tt}(\theta) \frac{d\,F(\tilde z)}{d\tilde z}&{\cal O}(\zeta^{-2})&\zeta h_{t\theta}(\theta) F(\tilde z)&
\zeta h_{t\phi}(\theta) \frac{d\,F(\tilde z)}{d\tilde z}&\zeta h_{t\tilde z}(\theta)\frac{d\,F(\tilde z)}{d\tilde z}\\
{\cal O}(\zeta^{-2})&
{\cal O}(\zeta^{-3})&
{\cal O}(\zeta^{-2})&{\cal O}(\zeta^{-2})
&\zeta^{-1}h_{\zeta \tilde z}(\theta)\frac{d^2\,F(\tilde z)}{d\tilde z^2}\\
\zeta h_{t\theta}(\theta)F(\tilde z)
&{\cal O}(\zeta^{-2})
&{\cal O}(\zeta^{-2})
&h_{\theta\phi}(\theta)F(\tilde z)&h_{\theta \tilde z}(\theta)F(\tilde z)\\
\zeta h_{t\phi}(\theta)\frac{d\,F(\tilde z)}{d\tilde z}&
{\cal O}(\zeta^{-2})
&h_{\theta\phi}(\theta)F(\tilde z)&
{\cal O}(\zeta^{-2})
&h_{\phi \tilde z}(\theta)\frac{d\,F(\tilde z)}{d\tilde z}\\
\zeta h_{t\tilde z}(\theta)\frac{d\,F(\tilde z)}{d\tilde z}&\zeta^{-1}h_{\zeta \tilde z}(\theta)\frac{d^2\,F(\tilde z)}{d\tilde z^2}&h_{\theta \tilde z}(\theta)F(\tilde z)&h_{\phi \tilde z}\frac{d\,F(\tilde z)}{d\tilde z} &h_{\tilde z\tilde z}(\theta)\frac{d\, F(\tilde z)}{d\tilde z}
\end{pmatrix}.\nonumber\\ \label{metric_fluct}
\eea
The allowed metric fluctuations given in (\ref{metric_fluct}) are such that the constraint equations $G^0_\mu=g^{0\nu}G_{\mu\nu}=0$, where $G_{\mu\nu}$ is the Einstein tensor, are satisfied. This restricts the number of independent functions in the metric fluctuations:
\bea
&&h_{\theta \tilde z}(\theta)=-\frac{1}{\sqrt 2}\frac{q}{D} h_{t\theta}(\theta)\nonumber\\
&&h_{tt}(\theta)=\frac{D^4}{q^4}h_{tt}=const \nonumber\\
&&h_{\zeta \tilde z}(\theta)=\frac{1}{2}h_{tt}\nonumber\\
&&h_{t\phi}(\theta)=-\frac{D^2}{q^2}h_{tt} \cos\theta\nonumber\\
&&h_{\tilde z\tilde z}(\theta)=-\sqrt{2} \frac{q}{D}h_{t\tilde z}(\theta)-\frac{D^2}{q^2}
h_{tt}\nonumber\\
&&h_{\phi \tilde z}(\theta)=-\frac{1}{2}\bigg(\sqrt{2}\frac{D}{q} h_{tt}\cos\theta
+2 \frac{q^2}{D^2} h_{t\tilde z}(\theta)\cos\theta-\frac{q^2}{D^2} \frac{d\, h_{t\tilde z}(\theta)}{d\theta}\sin\theta
+\frac{q^2}{D^2} h_{t\theta}(\theta)\sin\theta\bigg)\nonumber\\
&& h_{\theta \phi}(\theta)=-\frac{1}{2 \sin\theta \cos\theta}
\frac{D}{q} \bigg(\frac{q^3}{D^3} h_{t\theta}(\theta)\sin\theta\cos^2\theta
+2\sqrt{2}h_{tt}\cos\theta +2\frac{q}{D} h_{\phi \tilde z}(\theta)\cos^2\theta
\nonumber\\&&\;\;\;\;\;\;\;\;\;\;\;-
\frac{q^3}{D^3}\frac{d\, h_{t\tilde z}(\theta)}{d\theta}\sin\theta\cos^2\theta-
2\frac{q}{D}\frac{d\,h_{\phi \tilde z}(\theta)}{d\theta}\cos\theta\sin\theta
-\frac{q^3}{D^3}\frac{d\, h_{t\tilde z}(\theta)}{d\theta}\sin\theta\nonumber\\
&&\;\;\;\;\;\;\;\;\;\;\;+4 \frac{q^3}{D^3} h_{t\tilde z}(\theta)\cos\theta+
\frac{q^3}{D^3} h_{t\theta}(\theta)\sin\theta +2 \frac{q}{D}
h_{\phi \tilde z}(\theta)\bigg).
\eea
To leading order in $1/\zeta$, the most general diffeomorphism generator which preserves these boundary conditions is
\bea
&&\xi_n\equiv \xi_n^\mu \partial_\mu= \exp(-\frac{2\pi i n \tilde z}{L_{\tilde z}})\bigg(-ia n \zeta \partial_\zeta +f_{\phi}(\theta)\partial_\phi-\frac{L_{\tilde z}}{2\pi} \partial_{\tilde z}\bigg)\nonumber\\ \label{xi}
&&a=\frac {D^2 h_{tt} \pi}{L_{\tilde z} q^2}\nonumber\\
&&f_\phi(\theta)=-\frac{1}{2\sqrt{2}\cos\theta}\bigg(\sqrt{2}h_{t\tilde z}(\theta)+\frac{D}{q}\frac{L_{\tilde z}}{\pi}(a-1)\bigg),
\eea
where
\be
L_{\tilde z}= 4\pi \sqrt{2}\frac{q}{D}
\ee
is the period of the diagonalized $\tilde z$ coordinate in (\ref{6ddiagonal}).

We would like to highlight that, in contrast to the structure of the boundary conditions of the NENH 4d Kerr black hole and the diffeomorphism boundary-preserving generator of that case, here we have a free parameter $h_{tt}$ in the metric fluctuations, and correspondingly, a free parameter $a$ in the diffeomorphism $\xi_n$ given in (\ref{xi}). The constraint equations are satisfied with arbitrary $h_{tt}, h_{t\theta}(\theta), h_{t\tilde z}(\theta)$.
The relative normalization of the various terms in the diffeomorphism generator (\ref{xi}) was chosen such that its Lie algebra is a $Diff(S^1)$ algebra.

Let us now substitute (\ref{xi}) into (\ref{silva0}) and (\ref{bb0}). By casting the central term of Silva's conserved charge algebra into the standard Virasoro form we arrive at the following central charge expression\footnote
{
We have analyzed what happens if a different linear combination of the torus
$U(1)\times U(1)$ isometries is chosen, instead of $\partial_{\tilde z}$, to construct the vector field $\xi_n$ and the corresponding boundary conditions. The central charge dependence on the free parameter $a$ comes out the same, for each of the two diffeomorphism charge algebras.
}:
\bea
&&c_{S}={3\sqrt{2}}\frac{q a}{L_{\tilde z} D}\frac{A_{horizon;6d}}{G_6}\nonumber\\
&&\;\;\;\;\;\;=\frac{3 a A_{horizon;6d}}{4\pi G_6},
\eea
whereas from the BBC central term we obtain the central charge
\bea
&&c_{BBC}=\frac{3\sqrt{2}}{2} \frac{q a (3-a)}{L_{\tilde z} D}\frac{A_{horizon;6d}}{G_6}
\nonumber\\
&&\;\;\;\;\;\;=\frac{3 a (3-a) A_{horizon;6d}}{8\pi G_6},
\eea
where $A_{horizon;6d}=A_{horizon;4d}L_y L_z$ is the area of the horizon of the extremal 6d black hole.
The central charge does not depend on the free function $f_{\phi}$. On the other hand, $f_{\phi}$ will enter the expression of the zero mode ${\cal Q}_0$.

We note that for $$a=1$$ the two central charges agree, and the difference term (\ref{diffterm}) vanishes. It is for that same value $a=1$ that the conditions for the validity of the BBC algebra, in particular the second condition (\ref{condition}), are satisfied. 

We have also considered three other sets of diffeomorphisms analogous to that given in (\ref{xi}), but which depend solely on $t$, $\phi$, or $\tilde y$ respectively instead of $\tilde z$. Although each of these form a $Diff(S^1)$, we found that the Virasoro algebra resulting from either Silva's or the BBC prescription had vanishing central term and thus led to a vanishing entropy. We conclude that only the $\tilde z$-dependent diffeomorphisms are related to a non-trivial CFT dual. 

To obtain the extremal black hole entropy, we will use the thermal Cardy formula. The temperature of the dual chiral CFT $T_{\tilde z}$ is extracted from
\be
dS_{extremal}=\beta_{\tilde z} dN_{\tilde z}\equiv \frac{1}{T_{\tilde z}}dN_{\tilde z},
\ee
where $N_{\tilde z}$ is the quantized $U(1)$ charge associated with $\tilde z$-translations. For the NENH geometry (\ref{6ddiagonal}), the $U(1)$ charge is
\be
J_{\tilde z}=\frac{2\sqrt{2}\pi^2 q^3 (2+C^2-D^2)}{D^3 G_6}.
\ee
The quantized charge is obtained from the Bohr-Sommerfeld quantization formula
\be
N_{\tilde z}=\frac{1}{2\pi} L_{\tilde z} J_{\tilde z}.
\ee
Further using (\ref{S_extremal}) leads to
\be
T_{\tilde z} = \frac{1}{\pi}.
\ee

Finally, the thermal Cardy formula yields the Bekenstein-Hawking entropy
\be
S_{extremal}=\frac{\pi^2}3 cT = \frac{A_{horizon;6d}}{4G_6},
\ee
provided that $a=1$. The Bekenstein-Hawking formula for the entropy of the 4d extremal 2-charge black hole is immediately reproduced, since the 6d Newton's constant and the 4d Newton's constant are related by $L_y L_z/G_6 = 1/G_4$, and
the 4d black hole horizon area is $Area_{horizon;4d}= Area_{horizon;6d}/(L_y L_z)$.

\section*{Acknowledgments}

The authors are grateful to G. Comp\`{e}re for useful correspondence, and comments. 



\end{document}